\documentclass[12pt]{article}
\usepackage{amssymb,amsmath,graphicx}

\newcommand{\be}{\begin{equation}}
\newcommand{\ee}{\end{equation}}
\newcommand{\bea}{\begin{eqnarray}}
\newcommand{\eea}{\end{eqnarray}}
\newcommand{\beas}{\begin{eqnarray*}}
\newcommand{\eeas}{\end{eqnarray*}}

\newcommand{\phic}{\varphi}

\newcommand{\la}{\langle}
\newcommand{\ra}{\rangle}

\def\bs#1{\renewcommand{\baselinestretch}{#1}}

\begin{document}
\begin{titlepage}
\begin{center}
{\Large \bf On three dimensions as the preferred dimensionality of space via the Brandenberger-Vafa mechanism}

\vspace{6mm}

\renewcommand\thefootnote{\mbox{$\fnsymbol{footnote}$}}
Brian Greene${}^1$\footnote{greene@phys.columbia.edu},
Daniel Kabat${}^{1,2}$\footnote{daniel.kabat@lehman.cuny.edu} and
Stefanos Marnerides${}^1$\footnote{stefanos@phys.columbia.edu}

\vspace{4mm}

${}^1${\small \sl Institute for Strings, Cosmology and Astroparticle Physics} \\
{\small \sl and Department of Physics} \\
{\small \sl Columbia University, New York, NY 10027 USA}

\vspace{4mm}

${}^2${\small \sl Department of Physics and Astronomy} \\
{\small \sl Lehman College, City University of New York} \\
{\small \sl Bronx, NY 10468 USA}

\end{center}

\vspace{1cm}

\noindent
In previous work it was shown that, in accord with the
Brandenberger-Vafa mechanism, three is the maximum number of spatial
dimensions that can grow large cosmologically from an initial thermal
fluctuation. Here we complement that work by considering the
possibility of successive fluctuations.  Suppose an initial
fluctuation causes at least one dimension to grow, and suppose
successive fluctuations occur on timescales of order $\alpha'^{1/2}$.
If the string coupling is sufficiently large, we show that such
fluctuations are likely to push a three-dimensional subspace to large
volume where winding modes annihilate.  In this setting three is the
preferred number of large dimensions. Although encouraging, a more
careful study of the dynamics and statistics of fluctuations is
needed to assess the likelihood of our assumptions.

\end{titlepage}
\setcounter{footnote}{0}
\renewcommand\thefootnote{\mbox{\arabic{footnote}}}

\section{Introduction\label{sect:intro}}

The Brandenberger-Vafa mechanism \cite{brandvafa} is one of the few
proposals within string cosmology for a mechanism that yields
dynamical cosmological growth of three spatial dimensions.  The
idea that the most basic property of our universe could follow from
the dimensionality of fundamental strings is very appealing.  For
reviews see \cite{Battefeld:2005av,Brandenberger:2008nx,Brandenberger:2011et},
and for numerical simulations in support of the scenario see \cite{Sakellariadou:1995vk}.

A closer examination of string gas dynamics, however, reveals
certain obstacles.  To set the stage for
the present work, it was found in \cite{winds,Danos:2004jz} that -- assuming
validity of the lowest-order dilaton gravity equations of motion --
the dilaton can quickly roll to weak coupling, leaving the universe
trapped in the Hagedorn phase.  Moreover it was found that -- assuming
the string gas could be treated as homogeneous -- there was no sharp
dimension dependence in the string annihilation rate.  So given these
assumptions, the dynamics did not single out three dimensions as
special.

By relaxing the assumption of homogeneity, it was shown in \cite{US}
that the mechanism can indeed operate if one takes into account the
fact that when winding strings are dilute enough (their mean
separation is large compared to their characteristic quantum
thickness) they behave semi-classically and their annihilation rates
are highly suppressed in more than three large spatial dimensions. The
conclusion of \cite{US} was that if the string gas fluctuates out of
the Hagedorn regime to a radiation regime in $d$ effective spatial
dimensions, then any remaining winding modes are indeed dilute enough
that they generically freeze out for $d > 3$ and annihilate for $d =
3$.\footnote{This assumes the thermal fluctuation takes place while
the string coupling is still large enough for string interactions to
be effective.  It also raises a serious issue, noted but not addressed in
\cite{US} or the present work, of whether a dilute string gas can
evolve to give a homogeneous universe.} Therefore three
dimensions is the \textit{maximum} number of spatial dimensions that
can grow large cosmologically due to an initial thermal fluctuation.

In the present work we would like to examine whether three dimensions
is also the \textit{minimum} number of large dimensions that can
result from a thermal fluctuation. A priori it seems easier for one or
two spatial dimensions to decompactify via the BV mechanism, since it
seems more likely that winding modes will find each other in this
lower dimensional subspace. If the BV mechanism is indeed the reason
the universe has three cosmologically large dimensions, we should
address the question of why we don't observe an effectively
lower-dimensional universe.

Answers to this question often turn to anthropic arguments. However in
the original proposal of Brandenberger and Vafa, the authors argued
that successive thermal fluctuations in the Hagedorn phase would
eventually cause the maximum number of dimensions to decompactify.
Here we would like to examine this possibility in more detail.  We postulate
an initial fluctuation that causes one or two dimensions to grow
large, and investigate the likelihood of subsequent fluctuations
causing a total of three dimensions to decompactify.  Thus, in
contrast to our previous work \cite{US}, we allow for multiple thermal
fluctuations.  We also relax the assumption of isotropy made in
\cite{US}.  Besides addressing the question of how many dimensions decompactify,
our results will shed light on rapid thermal fluctuations as a possible mechanism for
overcoming the obstacle to the BV mechanism pointed out in \cite{winds,Danos:2004jz}, namely
the dilaton rolling to weak coupling.

Essential to our discussion is the fact that, while there might be a
window of opportunity for three dimensions to decompactify once one or
two dimensions have grown large, there can be no such window for more
than three dimensions.  That is, anisotropic expansion of one or two
dimensions will never favor the eventual decompactification of more
than three dimensions. To see this, consider the case of $d>3$
isotropic large dimensions, where the equilibrium state is radiation
in $d$ dimensions and the winding modes want to annihilate. The annihilation rate
of these winding modes depends on the size $R$ of the large dimensions
in two ways \cite{US}. First, there is an enhancement $\sim R^2$ reflecting the
fact that longer strings are more likely to annihilate.\footnote{In
the T-dual picture these strings carry more momentum and are more
likely to interact.} Second, with $\Delta
x\sim \sqrt{\log R}$ the characteristic quantum thickness of a string
of length $R$, the amplitude has an impact parameter suppression $\sim
\exp(-({b}/{\Delta x})^2)$ in the $d-3$ large directions
transverse to string collisions. This exponential suppression was put
forward in \cite{US} as essential to the BV mechanism, in the sense that
winding modes will freeze out if $d>3$. Now imagine we let some
of these dimensions shrink while others grow.  We do this to effect
anisotropy, however we preserve the total volume (and energy) so
the equilibrium state does not change. This will only further (and
exponentially) suppress the annihilation of winding modes around the
smaller dimensions, by increasing the impact parameters in the larger
dimensions.  So strings winding the smaller dimensions are even less likely to annihilate than in the isotropic case
studied in \cite{US}.  Since the isotropic case already singled out $d \leq 3$, we conclude that only $d \leq 3$ dimensions can decompactify as
a result of anisotropic fluctuations.

Hence studying anisotropic fluctuations within a three-dimensional
subspace accommodates the relevant cases for decompactification.
Given an initial thermal fluctuation that causes some number of
dimensions to grow, our goal is to study the likelihood of subsequent
fluctuations causing three dimensions to decompactify. To keep the
investigation tractable we will only consider one degree of
anisotropy, namely between one large and two small dimensions, or
between one small and two large dimensions.

An outline of this paper is as follows.  In section
\ref{sect:dynamics} we set up the dynamical equations and discuss the possible
equilibrium phases of the system.  In section \ref{sect:fluctuations}
we give our procedure for choosing initial conditions and sampling
thermal fluctuations.  In section \ref{sect:results} we present our
numerical results, and in section \ref{sect:discussion} we discuss
their implications for the Brandenberger-Vafa mechanism.

\section{Dynamics\label{sect:dynamics}}

The general setup will follow the lines of \cite{US}, which the reader
may refer to for more details. The difference is that here we are
considering two (logarithmic) scale factors, $\nu$ and $\lambda$,
respectively for the $d_1<3$ dimensions initially unwound and growing,
and for the $d_2=3-d_1$ dimensions subsequently expanding after a
fluctuation, but wrapped with winding modes.

We work in type IIA string theory, on a flat toroidal background
with metric (in $\alpha'=1$ units)
\begin{equation}
ds^2=-dt^2+e^{2\nu (t)}\sum_{i=1}^{d_1}dx_i^2+e^{2\lambda(t)}\sum_{i=d_1+1}^3dx_i^2\hspace{8mm}0\leq x_i\leq 2\pi
\end{equation}
All other dimensions are held fixed at the self-dual radius. We also have
the homogeneous shifted dilaton field $\varphi(t) = 2\phi(t) -
d_1\nu(t) - d_2\lambda(t)$. When the metric and dilaton are coupled to
matter, the equations of motion are
\begin{equation}\label{eqmots1}
\begin{aligned}\ddot{\varphi}&=\frac{1}{2}(\dot{\varphi}^2+d_1\dot{\nu}^2+d_2\dot{\lambda}^2)\\
\ddot{\nu}&=\dot{\varphi}\dot{\nu}+\frac{1}{8\pi^2}e^{\varphi}P_{\nu}\\
\ddot{\lambda}&=\dot{\varphi}\dot{\lambda}+\frac{1}{8\pi^2}e^{\varphi}P_{\lambda}
\end{aligned}
\end{equation}
Here $P_{\nu},\,P_{\lambda}$ are the pressures along the $d_1,\,d_2$
dimensions respectively. Note that we do not consider a potential for
the dilaton.  There is also the Hamiltonian constraint (or Friedmann
equation)
\begin{equation}\label{ham1}
E=(2\pi)^2e^{-\varphi}(\dot{\varphi}^2-d_1\dot{\nu}^2-d_2\dot{\lambda}^2)
\end{equation}
where $E$ is the total energy in matter. We take matter to consist of
\begin{itemize}
\item
Winding modes, with $W$ denoting the winding number of strings wound with positive orientation.  For simplicity we assume this winding number
is carried by $W$ individual strings, each having a single unit of positive winding.  This is a conservative assumption, since taking multiply-wound strings
into account would lead to a larger annihilation rate.\footnote{See for example (30) in \cite{winds}.}  Also note that since we are working in a compact space,
the net winding number must vanish, which means there are an equal number of strings wound with the opposite orientation. Since the $d_1$
dimensions are assumed unwound to begin with, the winding modes only
wrap the dimensions $d_2$. These winding modes evolve according to
\begin{equation}\label{boltzW}
\dot{W}=-\Gamma_W(W^2-\langle W\rangle^2)
\end{equation}
Here $\langle W \rangle$ denotes the equilibrium average winding and
$\Gamma_W$ is an interaction rate we specify below.  The contribution
to the energy from winding and anti-winding modes is $E = 2 d_2 W
e^{\lambda}$ and their contribution to the (off-equilibrium) pressure
is $P_\lambda =-2We^{\lambda}$.
\item
Radiation, or pure Kaluza-Klein modes, with $K_{\nu}$ and
$K_{\lambda}$ denoting the momentum numbers of strings with positive momentum along the $d_1$ and
$d_2$ dimensions respectively. We assume this momentum is carried by individual
strings that each have one unit of positive Kaluza-Klein momentum; the net momentum vanishes so there must be an equal number of strings with negative momentum. These momentum modes evolve according to
\begin{equation}\label{boltzK}\begin{aligned}
\dot{K}_{\nu}&=-\Gamma_{K_{\nu}} (K_{\nu}^2-\langle K_{\nu}\rangle^2)\\
\dot{K}_{\lambda}&=-\Gamma_{K_{\lambda}} (K_{\lambda}^2-\langle K_{\lambda}\rangle^2)
\end{aligned}\end{equation}
The energy in these modes is
$E=2d_1K_{\nu}e^{-\nu}+2d_2K_{\lambda}e^{-\lambda}$ and their
contributions to the pressures are $P_{\nu} = 2K_{\nu}e^{-\nu}$,
$P_{\lambda} = 2K_{\lambda}e^{-\lambda}$.
\item
String oscillator modes which we model as pressureless matter.  
\end{itemize}

\subsection{Equilibrium phases}

In this section we review the possible equilibrium phases of a string gas.  This information will be important in the sequel,
when we study fluctuations and the approach to equilibrium.  A point of terminology: in the remainder of this paper,
when we refer to the universe as being in one of these possible phases, {\em we do not mean that the universe is actually
in thermal equilibrium}.  Rather we are using these names as a convenient shorthand to indicate what the equilibrium
state of the universe would be, given its energy density.

In the $d$-dimensional isotropic case the string gas has two possible equilibrium
thermodynamic phases. In the Hagedorn phase massive and massless modes
are in thermal equilibrium, the pressure vanishes and the energy in
matter is conserved. At lower energy densities the equilibrium state of the
universe is radiation dominated.  This occurs when the energy density
in $d$ dimensions satisfies
\begin{equation}\label{denscon}
\rho_d=\frac{E}{V_d}<\rho_H=c_dT_H^{d+1}
\end{equation}
with $T_H$ the Hagedorn temperature and $c_d$ the Stefan-Boltzmann
constant in $d$ dimensions. This can be alternatively expressed in
terms of temperatures.  In a volume $V_d$ with energy $E$, a radiation
gas has temperature $T_d = (\frac{E}{c_d V_d})^{\frac{1}{d+1}}$. The
condition (\ref{denscon}) can be then expressed as
\begin{equation}\label{Temp1}
T_d<T_H
\end{equation}
That is, the universe is radiation-dominated when the would-be
radiation temperature falls below the Hagedorn temperature.

In the anisotropic case there is an additional
possibility. Recall that we have three large dimensions, with $d_1$ of
them larger than the remaining $d_2=3-d_1$.  Besides the Hagedorn
phase, the system could be found in either a 3-dimensional or a
$d_1$-dimensional radiation phase.  To fix the equilibrium phase of
the universe we generalize the condition (\ref{Temp1}). Given the
energy of the system and the 3-dimensional and $d_1$-dimensional
volumes, the lower of the two temperatures
$T_{d_1}=(\frac{E}{c_{d_1}V_{d_1}})^{\frac{1}{d_1+1}}$ and
$T_3=(\frac{E}{c_3V_3})^{\frac{1}{4}}$ determines the equilibrium
phase. If neither of these temperatures is lower than $T_H$ then the
system is in the Hagedorn phase. For brevity, we will refer to these
equilibrium phases as $\mathcal{R}_{d_1}$, $\mathcal{R}_3$ and $\mathcal{H}$.

In equilibrium in $\mathcal{R}_{d_1}$ the energy is entirely carried by the massless
Kaluza-Klein modes $K_{\nu}$. Treating these modes as a collection of one-dimensional gasses, the equilibrium values are \cite{winds}
\begin{equation}
\label{Rd1W}
\mathcal{R}_{d_1}:\hspace{3mm}\begin{aligned}
&\langle K_{\nu}\rangle =\frac{E}{2d_1}e^{\nu},\\ &\langle
K_{\lambda}\rangle=\langle W\rangle=0\end{aligned}
\end{equation}
These values account for all of the available energy, see below (\ref{boltzK}).
They correspond to pressures
\begin{equation}
\label{Rd1pressure}
\mathcal{R}_{d_1}:\hspace{3mm}
\langle P_{\nu} \rangle =\frac{E}{d_1},\hspace{2mm}
\langle P_{\lambda} \rangle = 0
\end{equation}
In equilibrium in $\mathcal{R}_3$ the energy is split between $K_{\nu}$ and
$K_{\lambda}$, $E=2d_1K_{\nu}e^{-\nu} +
2d_2K_{\lambda}e^{-\lambda}$. Recalling that $d=3=d_1+d_2$ the equilibrium values are
\begin{equation}
\label{R3W}
\mathcal{R}_3:\hspace{3mm}\begin{aligned}
&\langle K_{\nu}\rangle =\frac{E}{6}e^{\nu},\\
&\langle K_{\lambda}\rangle =\frac{E}{6}e^{\lambda},\hspace{2mm}\langle W\rangle=0\end{aligned}
\end{equation}
corresponding to pressures
\begin{equation}
\label{R3pressure}
\mathcal{R}_3:\hspace{3mm}
\langle P_{\nu} \rangle = \langle P_{\lambda} \rangle = \frac{E}{3}
\end{equation}
Finally in the Hagedorn phase the equilibrium values are \cite{winds}
\begin{equation}
\mathcal{H}:\hspace{3mm}\begin{aligned}
&\langle K_{\nu}\rangle =\frac{1}{12}\sqrt{\frac{E}{\pi}}e^{\nu},\\
&\langle K_{\lambda}\rangle
=\frac{1}{12}\sqrt{\frac{E}{\pi}}e^{\lambda},\hspace{2mm} \langle
W\rangle=\frac{1}{12}\sqrt{\frac{E}{\pi}}e^{-\lambda}\end{aligned}
\end{equation}
with pressures (due to the lack of winding around $d_1$)
\begin{equation}
\mathcal{H}:\hspace{3mm}
\langle P_{\nu} \rangle =\frac{1}{6}\sqrt{\frac{E}{\pi}},\hspace{2mm}
\langle P_{\lambda} \rangle = 0
\end{equation}
We also recall the off-equilibrium interaction rates \cite{winds}. In the radiation phases, for strings with a single unit of momentum or winding, the rates are
\begin{equation}
\label{rates}
\Gamma_{K_{\nu}}=\frac{1}{\pi}e^{-2\nu+\varphi},\hspace{3mm}
\Gamma_{K_{\lambda}}=\frac{1}{\pi}e^{-2\lambda+\varphi},\hspace{3mm}
\Gamma_{W}=\frac{1}{\pi}e^{2\lambda+\varphi}.
\end{equation}
In the Hagedorn phase these rates get multiplied by a factor
$16 \pi E / 9$ reflecting the enhancement due to highly excited oscillator
modes.\footnote{This follows from (19), (28), (29) in \cite{winds}.}

\section{Fluctuations, initial conditions and procedure\label{sect:fluctuations}}

We are interested in the likelihood of decompactifying three dimensions as a result of successive
fluctuations.  To this end we consider the following scenario.  The first fluctuation occurs at time $t=0$ and takes the universe to a
state in which $d_1<3$ dimensions are expanding and free from winding. The initial size of these
dimensions is $\nu_0 > 0$ and their initial expansion rate is
$\dot{\nu}_0 > 0$. For simplicity we treat all other dimensions as fixed in size, with vanishing
expansion rates. Then at time $t_f$ some number of dimensions $d_2=3-d_1$
fluctuate to a size $\lambda(t_f)$ and begin expanding at a rate
$\dot{\lambda}(t_f) > 0$.  We assume that $ 0 < \lambda(t_f) <
\nu(t_f)$.  This models an energy fluctuation with energy flowing from
matter to gravitation. We are interested in the likelihood that all three dimensions will decompactify,
that is, that the winding modes
wrapping the $d_2$ dimensions will annihilate and that all three dimensions will begin to expand.  We will study this as a function of
the initial conditions, the fluctuation time $t_f$ and the
\textit{anisotropy parameter}
\begin{equation}\label{anis}
r\equiv\frac{R_{\lambda}}{R_{\nu}}(t_f)=e^{\lambda(t_f)-\nu(t_f)}
\end{equation}

Depending on the energy density, the universe could start out at $t = 0$ with any one of
the three possible equilibrium phases $\mathcal{R}_3$, $\mathcal{R}_{d_1}$, or
$\mathcal{H}$.  If the universe starts with $\mathcal{R}_3$ as its equilibrium phase
then there is no need for a subsequent fluctuation: there will be few or no winding strings present -- from
(\ref{R3W}) we have $\langle W \rangle = 0$ -- so the initial conditions will most likely
directly lead to three-dimensional decompactification.  Another possibility is that
at time $t=0$ the universe could start with
$\mathcal{R}_{d_1}$ as its equilibrium phase. In this case it is quite possible that the strings winding around the $d_2$ dimensions
will eventually annihilate.  This would allow the universe to relax to equilibrium in the phase $\mathcal{R}_{d_1}$.  But
from (\ref{Rd1W}), (\ref{Rd1pressure}) note that in equilibrium in $\mathcal{R}_{d_1}$ we have $\langle K_{\lambda} \rangle = \langle W \rangle = 0$, which means the pressure in the
$d_2$ dimensions vanishes.  So even though the $d_2$ dimensions can shed their winding, it seems unlikely
that these dimensions will begin to expand.  Instead we expect them to remain small and unwound.  This leads us to
make a conservative assumption, that if the universe finds itself with $\mathcal{R}_{d_1}$ as its equilibrium phase it
will never become effectively three dimensional.    This assumption could be
lifted through a more careful study of fluctuations,
which might show some probability of decompactification even starting from $\mathcal{R}_{d_1}$. As we do not attempt such a
study here, we employ the conservative assumption that a universe which begins in $\mathcal{R}_{d_1}$
will never end up with three large dimensions.

Thus in order to obtain
three large dimensions from successive fluctuations, \textit{at time $t=0$ the system should find
itself with $\mathcal{H}$ as its equilibrium phase}. Given $\nu_0$, this puts a lower bound on
the energy at $t=0$,
\begin{equation}
\label{conr2} E\geq c_{d_1}V_{d_1}T_H^{d_1+1}
\end{equation}
At fixed volume, this essentially constrains the initial value of the
dilaton as we will see shortly.

One might worry that starting in $\mathcal{H}$ at $t=0$ leads to a
further restriction when we require that the $d_1$ dimensions are
initially unwound.  The issue is that in the Hagedorn phase the
equilibrium winding number does not vanish, unless the ratio
$E/e^{\nu}$ is small enough that no winding is energetically allowed. In
practice this is not a concern.  As we will see below, in order for the eventual
decompactification of three dimensions to take place, the initial
value of $\nu_0$ must be large enough, and the initial energy low
enough, that initial conditions which allow decompactification of three dimensions are
consistent with the condition that the $d_1$ dimensions start
out unwound. Another way to see this is to assume the contrary as
follows. Suppose we start deep in the Hagedorn phase where the $d_1$
dimensions are wound. Then when the $d_2$ dimensions fluctuate, it is
unlikely that they will be able to push the system out of
$\mathcal{H}$ and into $\mathcal{R}_3$.  Thus if the $d_1$ dimensions
start out wound, fluctuations can occur but will most likely just
leave the universe trapped in the Hagedorn phase.

At this point we review our procedure for fixing initial conditions.
The maximum value of $|\dot{\varphi}|$ consistent with the
supergravity approximation is $|\dot{\varphi}| = 1$. Orienting time so
the universe is rolling to weak coupling, we set
$\dot{\varphi}_0=-1$. This defines our initial time slice; the
equations of motion then guarantee that $|\dot{\varphi}(t)|<1$ and
$\dot{\varphi}(t)<0$ for all $t$.  To explore the dependence on
$\varphi_0$ we will scan over a range of values specified below,
consistent with the universe starting in the Hagedorn phase. Given
$\dot{\varphi}_0$ and $\varphi_0$, the expansion rate $\dot{\nu}_0$ is
chosen at random.  The total energy in the universe vanishes by the Hamiltonian constraint, so we assume the relevant probability distribution for $\dot{\nu}_0$ is given by the microcanonical ensemble.  This distribution turns out to have a Gaussian form.  To see this note that
\begin{equation}
\label{nugauss}
\text{prob}(\dot{\nu}_0) \sim \text{exp}(S) = \text{exp}(E/T_H) \propto\text{exp}\bigl[-\bigl(\frac{4\pi^2d_1e^{-\varphi_0}}{T_H}\bigr)\dot{\nu_0}^2\bigr]
\end{equation}
where we used the Hagedorn equation of state to determine the entropy, $S = E / T_H$, and we used
the Hamiltonian constraint (\ref{ham1}) to determine the energy available in matter.
Finally we make a choice for $\nu_0$, which controls the size of the
dimensions $d_1$.  In practice we will test two values, $\nu_0 = 3$
and $\nu_0 = 5$.  Note that via the constraint (\ref{ham1}), the
quantities $\dot{\varphi}_0$, $\varphi_0$, $\dot{\nu}_0$ are sufficient
to determine the energy in matter at $t=0$. Given the energy in matter
and a choice for $\nu_0$, the initial Kaluza-Klein momentum in the large dimensions
$K_{\nu}(t = 0)$ is set to its equilibrium value.

We then evolve the system to time $t_f$ using the equations of motion
(\ref{eqmots1}), (\ref{boltzW}), (\ref{boltzK}). During this time we hold $\lambda = 0$ fixed at the self-dual radius.
At time $t_f$ we
effect a fluctuation for the dimensions $d_2$. To model a fluctuation
we draw energy from the bath of heavy oscillator modes and
re-distribute that energy to the other matter and metric modes, i.e.\
to the expansion rates. We specify the size of the fluctuation, that
is the value of $\lambda(t_f)$, using the anisotropy parameter $r$
defined in (\ref{anis}).  In practice we will scan over the range $0.1
< r <1$. Then using $\dot{\varphi}(t_f)$, $\varphi(t_f)$ and
$\dot{\nu}(t_f)$ we evaluate the energy in matter for
$\dot{\lambda}=0$ and tentatively determine the equilibrium phase of
the universe.  This allows us to choose the expansion rate
$\dot{\lambda}(t_f)$ from a Gaussian probability distribution,
where the width of the Gaussian is determined as follows.  In the Hagedorn phase
we have a result similar to (\ref{nugauss}),
\begin{equation}
\text{prob}(\dot{\lambda}) \sim \text{exp}(S) = \text{exp}(E/T_H) \propto\text{exp}\bigl[-\bigl(\frac{4\pi^2d_2e^{-\varphi}}{T_H}\bigr)\dot{\lambda}^2\bigr]
\end{equation}
and we read off the variance
\begin{equation}
\sigma_{\mathcal{H}}^2 = \frac{T_He^{\varphi}}{8\pi^2d_2}
\end{equation}
In the radiation phase
$\mathcal{R}_3$, on the other hand, the matter entropy is
\begin{equation}
S = {d + 1 \over d} c_d V_d T^d = {d + 1 \over d} (c_d V_d)^{1 \over d + 1} E^{d \over d+1}
\end{equation}
Using (\ref{ham1}) to determine the available energy and expanding in powers of $\dot{\lambda}$ we have
\begin{equation}
\text{prob}(\dot{\lambda}) \sim \text{exp}(S) \propto\text{exp}\bigl[-\dot{\lambda}^2/2\sigma_{\mathcal{R}_3}^2\bigr]
\end{equation}
where the variance is
\begin{equation}
\sigma_{\mathcal{R}_3}^2 = {1 \over 2 d_2} \left({\dot{\varphi}^2 \over c_d V_d}\right)^{1 \over d+1} \left({e^\varphi \over 4 \pi^2}\right)^{d \over d+1}
\end{equation}
Once $\dot{\lambda}(t_f)$ is chosen we re-calculate the energy available to
matter. We still need to determine the values of $K_{\lambda}(t_f)$
and $W(t_f)$. These are chosen from a uniform distribution.  We
randomly select $K_{\lambda}(t_f)$ in the range
$(K_{\text{self-dual}},K_{\text{equilibrium}})$ and $W(t_f)$ in the
range $(\text{Max}\{0.5,\la W\ra\},W_{\text{self-dual}})$.  Note that
the values at the self-dual radius ($\lambda=0$) set a lower bound on
$K_{\lambda}$ and an upper bound on $W$.  Also the value $0.5$ is our
cutoff value for rounding down to no winding: we do not allow the
fluctuation to result in vanishing winding, as our goal is to test
whether the winding modes will annihilate as a consequence of
interactions.

In summary, the values $\varphi_0$, $\dot{\varphi}_0$, $\nu_0$,
$K_{\nu}(t=0)$ and $\lambda(t_f)$ (or $r$) are fixed and we randomly
choose $\dot{\nu}_0$, $\dot{\lambda}(t_f)$, $K_{\lambda}(t_f)$ and
$W(t_f)$ from the distributions given above. For $t > t_f$ we solve
the equations of motion until the winding modes either freeze out or
annihilate completely. In practice, when we numerically integrate the
equations of motion, the winding modes are considered annihilated when
$W<0.5$ and frozen when $\Gamma_WW<0.1\dot{\nu}$. (This is the relevant comparison of interaction rate and expansion rate, appropriate for the Boltzmann equation
(\ref{boltzW}).  The use of
$\dot{\nu}$ here reflects the fact that the winding modes collide in
the $d_1$ dimensions.)  This lies at the heart of our test of the BV
mechanism: the dimensions $d_1$ are unwound and free to expand, and we
test whether the winding modes wrapping the remaining $d_2$ dimensions
can stay in equilibrium as they compete with the expansion rate
$\dot{\nu}$ of the $d_1$ larger dimensions.

A note is in order regarding fluctuations of the scale factors. Given
the Hagedorn equation of state $S=E/T_H$ we assume the scale factors
can fluctuate randomly with a uniform distribution, since there is no
cost in entropy. This is true in the thermodynamic limit. However, the (equilibrium) entropy receives corrections at
large radii, given by \cite{deo1, borunda}
\begin{equation}
\Delta S = \text{log}\left[1-\sum_i\Gamma(2d_i)^{-1}(\eta_i
E)^{2d_i-1}e^{-\eta_iE}\right],\hspace{5mm}\eta_i=T_H^{-1}\left(1-\sqrt{1-\frac{1}{2R_i^2}}\,\right)
\end{equation}
for each set of $d_i$ dimensions with radius $R_i$. These corrections
are negligible when compared to the leading term, typically by many
orders of magnitude. The probability distribution resulting from this
correction as a function of $R$ at constant matter energy is very flat
in the Hagedorn phase.  There is a mild dip around $\eta_iE=2d_i-1$,
but this lies outside the Hagedorn phase except when $d_i=1$.  We
ignore these corrections and adopt a flat distribution for scale
factor fluctuations.

\section{Results\label{sect:results}}

For the scale factor $\nu_0$ associated with the initial size of the universe
we consider two possible values, $\nu_0 = 3$ and $\nu_0 = 5$. We will
see that the results only mildly depend on the choice of $\nu_0$.  As
values of $\nu_0$ larger than $5$ are in a sense ``too large,'' since
they correspond to radii far larger than the string scale, we trust
that these two values provide a good test of the Brandenberger-Vafa
mechanism.

Once $\nu_0$ is chosen, the condition (\ref{conr2}) that the system
starts in $\mathcal{H}$ at $t=0$ translates to the following condition
on the value of $\varphi_0$:
\be \label{dillim}
\varphi_0\lesssim-d_1\nu_0-\log [(2\pi)^{d_1-2}c_{d_1}T_H^{d_1+1}]
\ee
(We're using the fact that $\dot{\varphi}_0 = -1$, while the Hubble
rates in (\ref{ham1}) are much smaller than 1).  The above is not a
severe constraint since it is consistent with the prerequisite of weak
coupling, although it constrains the case $d_1=2$ more than
$d_1=1$ since it pushes us to weaker values for the coupling.  The
condition that the $d_1$ dimensions start out unwound, $\la W\ra <0.5$,
translates to
\be \label{dillim2}
\varphi_0\gtrsim -2\nu_0-\log [9/\pi]
\ee
Taken together, (\ref{dillim}) and (\ref{dillim2}) fix the range of
dilaton values we want to test. What happens outside this range? At
stronger values of the coupling, corresponding to less initial energy, the
system is found in $\mathcal{R}_{d_1}$ at $t = 0$.  As discussed in the paragraph
below (\ref{anis}), we make the conservative assumption that in this case additional dimensions will never decompactify,
even if they shed their winding.  On
the other hand at weaker values of the coupling the system is trapped forever in the
Hagedorn phase, with all dimensions wound.

For the fluctuation time $t_f$ we consider three possible values,
$t_f=1$, $10$, $100$ in string units ($\alpha '=1$). This turns out to
be the most decisive parameter, with $t_f=1$ giving the largest window
for decompactification of three dimensions.

To summarize the initial conditions, we fix
$\nu_0 \in \lbrace 3,5 \rbrace$ and $t_f \in \lbrace 1,10,100 \rbrace$.
We then scan over $0 < r < 1$ and the range of $\varphi_0$ dictated by
(\ref{dillim}) and (\ref{dillim2}). For each choice of initial
conditions we perform 100 integrations of the equations of motion to
sample values of $\dot{\nu}_0$, $\dot{\lambda}(t_f)$, $K_\lambda(t_f)$ and
$W(t_f)$ from the distributions mentioned above.

\begin{figure}[htp]
\centering
\includegraphics*[scale=0.8,viewport=90 130 550 750]{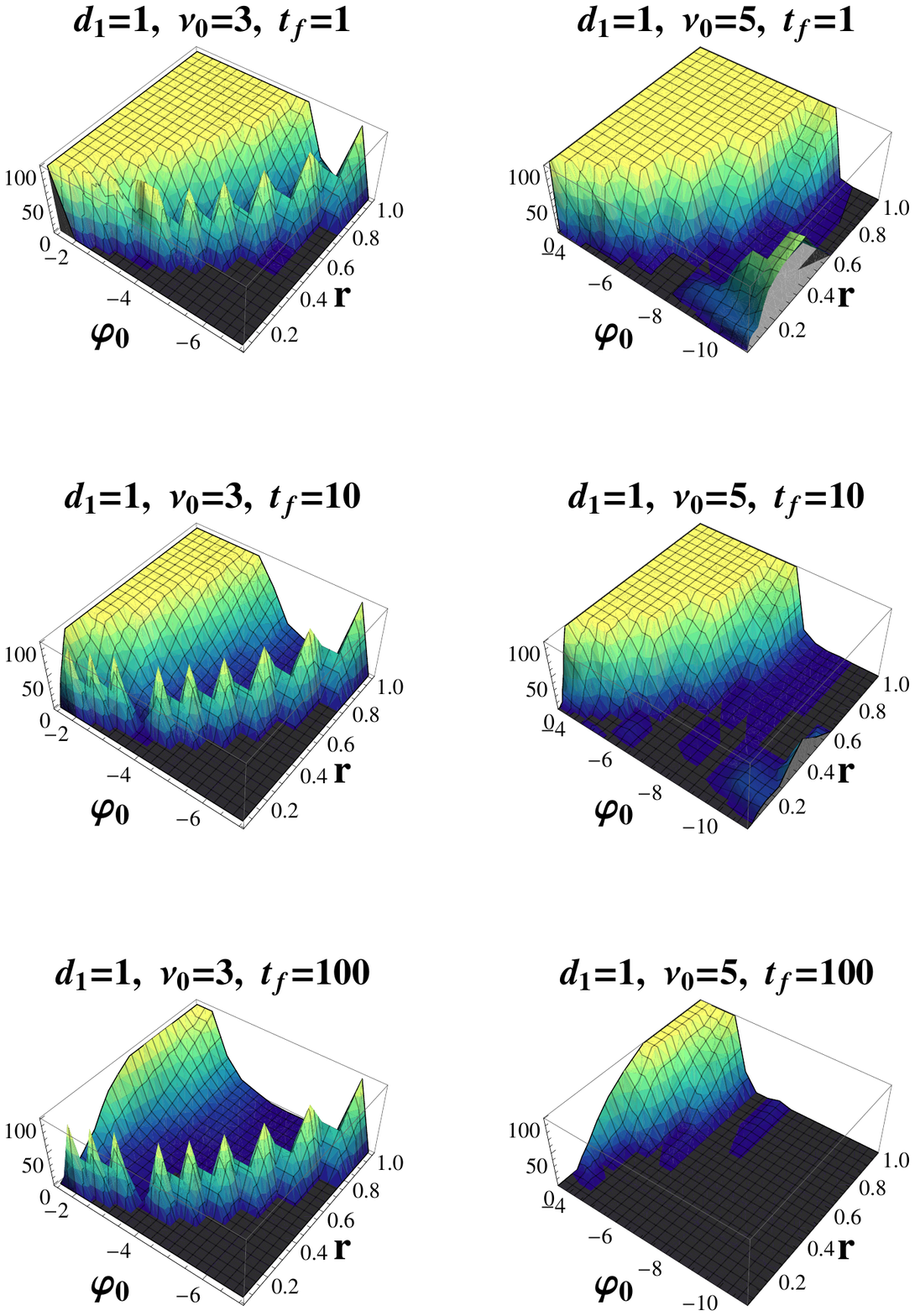}
\bs{1}
\caption{\small Number of cases decompactifying (out of 100) as a function of
$\phic_0$ and $r$, for $\nu_0=3,5$ and $t_f=1,10,100$ when $d_1=1$.}\bs{1.7}
\label{d113D}\end{figure}

\begin{figure}[htp]
\centering
\includegraphics*[scale=0.8,viewport=90 130 550 750]{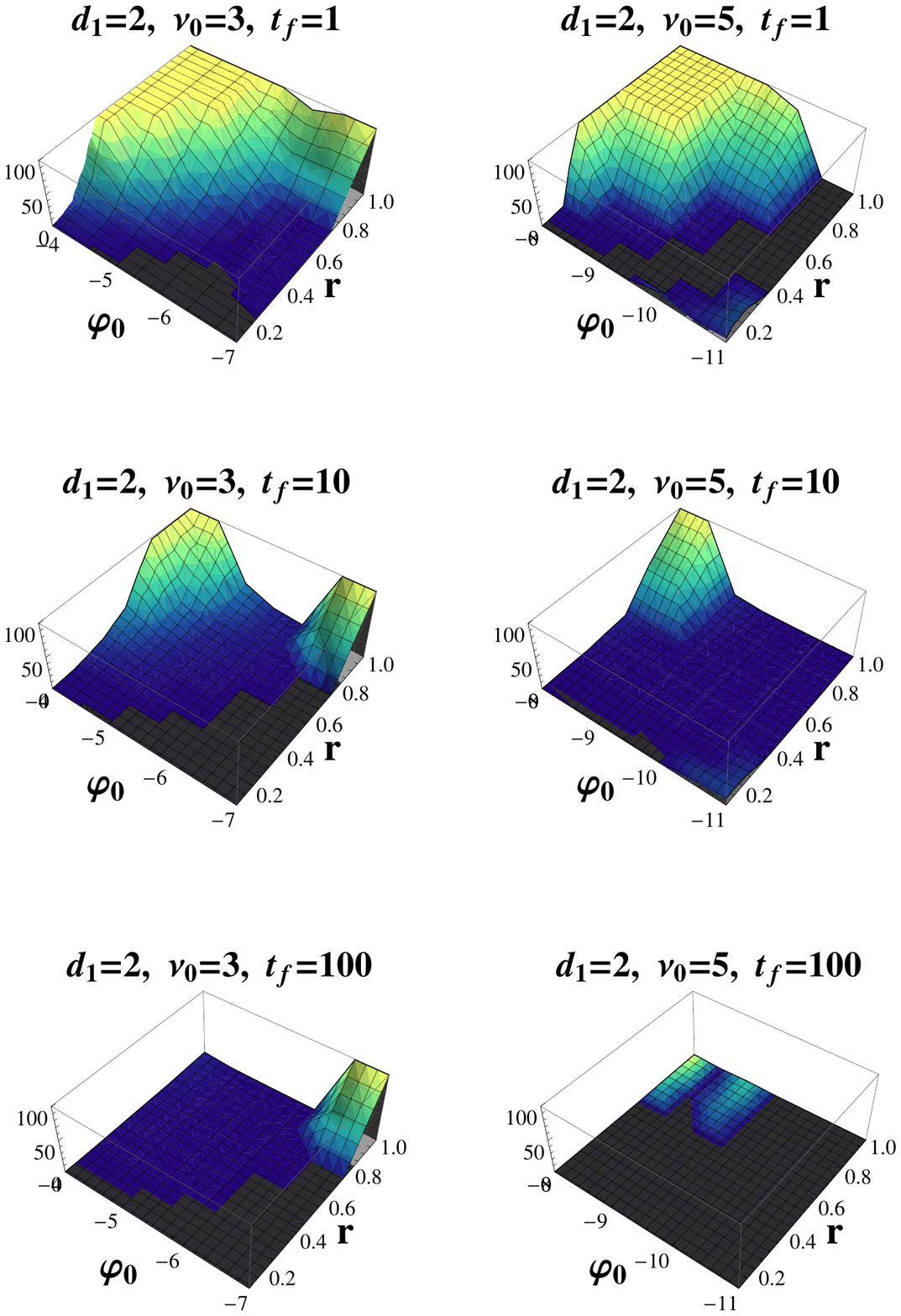}
\bs{1}
\caption{\small Number of cases decompactifying (out of 100) as a function of
$\phic_0$ and $r$, for $\nu_0=3,5$ and $t_f=1,10,100$ when $d_1=2$.}\bs{1.7}
\label{d123D}\end{figure}

Our results are shown in Figs.\ \ref{d113D} and \ref{d123D}, where we
plot the number of cases for which the winding modes annihilate as a
function of $\phic_0$ and $r$.  The dependence on $\varphi_0$ and $r$
is as expected, with strong coupling (large $\varphi_0$) and large
fluctuations (large $r$) more likely to decompactify.  The dependence
on $\nu_0$ is rather mild.  The constraints (\ref{dillim}),
(\ref{dillim2}) push us to weaker coupling as $\nu_0$ is increased,
but the enhancement in the annihilation amplitude $\sim e^{2
  \lambda(t_f)} = r^2 e^{2 \nu(t_f)}$ makes up for this suppression.
A striking feature is the dependence of the plots on $t_f$. The reason
decompactification becomes unlikely at large $t_f$ is the rolling of
the dilaton to weak coupling.  This suppresses the rates (\ref{rates})
and makes it impossible for wound strings to annihilate.  Note that this leads to a degeneracy,
that a large value for $t_f$ can be compensated by starting with a stronger coupling or a smaller
value of $\dot{\varphi}$.

\begin{figure}[htp]
\begin{center}
\vspace{-1cm}
\includegraphics*[scale=0.8,viewport=90 100 550 740]{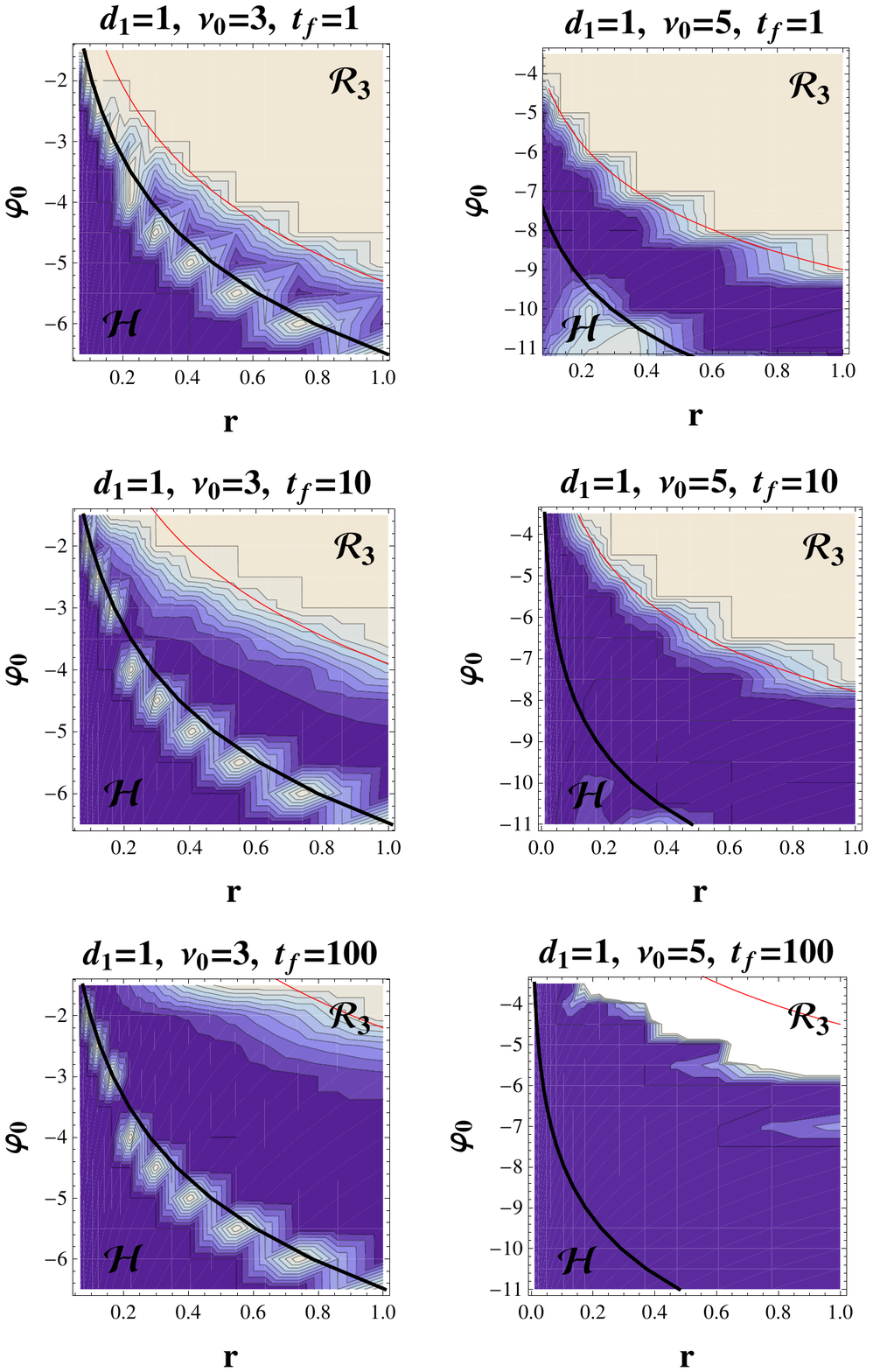}
\end{center}
\caption{\small Contours of constant probability of decompactification
  for $d_1 = 1$. The thick black line is the energy density
  separating the equilibrium radiation and equilibrium Hagedorn phases. The thin red line is
  the energy density below which more than $90\%$ of the cases
  decompactify.}\bs{1.7}
\label{d11cont}\end{figure}

\begin{figure}[htp]
\begin{center}
\includegraphics*[scale=0.8,viewport=90 140 550 765]{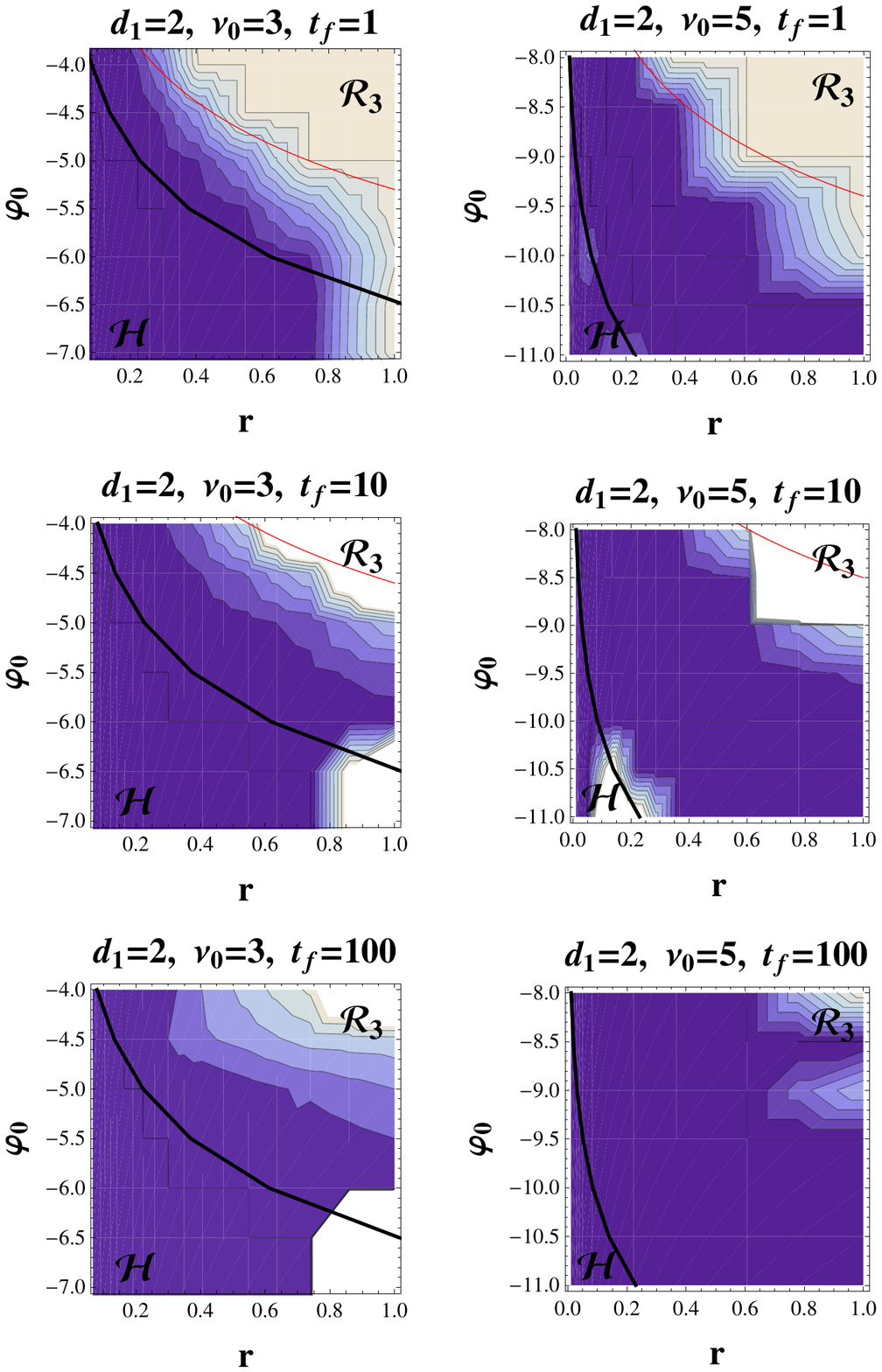}
\end{center}
\caption{\small Same as Figure \ref{d11cont} but for $d_1=2$.}\bs{1.7}
\label{d12cont}\end{figure}

In all cases, the decisive factor for decompactification is the energy
density of the universe.  This is shown in Figs.\ \ref{d11cont} and
\ref{d12cont}. On these graphs we show contour plots for the number of
cases decompactifying.  We also show the contour (thick black
line) corresponding to the energy
density that separates an equilibrium phase $\mathcal{H}$ from an equilibrium phase $\mathcal{R}_3$.  We also show the contour (thin red line) for the energy density below which more than $90\%$ of the cases
result in three dimensions decompactifying.

The main lesson is that a fluctuation is quite likely to make three dimensions decompactify provided the fluctuation is large enough to push the
equilibrium phase of the universe well into $\mathcal{R}_3$.  As in \cite{US}, a small fluctuation may leave the universe
trapped in $\mathcal{H}$, although what we mean by this here is that
the winding modes freeze out while the $d_1$ dimensions keep growing.
Note that there is a narrow region -- near the critical energy density separating
$\mathcal{H}$ and $\mathcal{R}_3$ -- where it is possible that $\la W\ra$ is non-zero after the fluctuation, but
drops to zero as the universe expands.  If this happens while interactions are still efficient the winding
modes will annihilate.  This is responsible for the spiky ridge seen in
Fig.\ \ref{d113D}.\footnote{The ridge lies on the $\mathcal{H}$ side of the divide between $\mathcal{H}$ and $\mathcal{R}_3$, because on the $\mathcal{H}$
side string interaction rates are enhanced by the factor $16 \pi E / 9$ discussed below (\ref{rates}).  A more refined treatment of the interaction rates of excited strings would
presumably smooth out the ridge to some extent.}
This discussion also makes it clear that the
contours of constant decompactification probability track the
contours of constant energy density.  To understand this, note that
energy density is the key parameter which determines whether the universe has Hagedorn
or radiation as its equilibrium phase. 

Since energy density is the most important parameter, in Table
\ref{denstab} we give the range of energy densities at time $t_f$ that
lead to three dimensions decompactifying.  We also express our results
as a function of $d_1$, $\nu_0$ and $t_f$ by giving the percentage of
cases that decompactify three dimensions in Table \ref{pertab}.

\begin{table}[h]
\begin{center}
\begin{tabular}{|c|c|c|c|} 
\hline 
\multicolumn{4}{|c|}{$d_1=1$}\\
\hline 
& $t_f=1$ & $t_f=10$ & $t_f=100$ \\ 
\hline
$\nu_0=3$& $(1.5\sim 65)\times 10^{-9}$ & $(1.5\sim 15)\times 10^{-9}$  &
$(1.5\sim 2.8)\times 10^{-9}$  \\
$\nu_0=5$& $(2.4\sim 640)\times 10^{-11}$ & $(2.4\sim 195)\times 10^{-11}$   &
$(2.4\sim 7.1)\times 10^{-11}$ \\ 
\hline 
\hline 
\multicolumn{4}{|c|}{$d_1=2$}\\
\hline 
& $t_f=1$ & $t_f=10$ & $t_f=100$ \\ 
\hline
$\nu_0=3$& $(1.4\sim 6.5)\times 10^{-8}$ & $(1.4\sim 3.1)\times 10^{-8}$  &
$0$  \\
$\nu_0=5$& $(2.3\sim 10)\times 10^{-9}$ & $(2.3\sim 3.9)\times 10^{-9}$   &
$0$ \\ 
\hline 
\end{tabular}
\end{center}\bs{1} \caption{\small The range of energy densities at time $t_f$ that lead to three
  dimensional decompactification.}\bs{1.7}\label{denstab}
\end{table}

\begin{table}[h]
\begin{center}
\begin{tabular}{|c|c|c|c|}
\hline
\multicolumn{4}{|c|}{$d_1=1$}\\
\hline
& $t_f=1$ & $t_f=10$ & $t_f=100$ \\
\hline
$\nu_0=3$& $82$ & $53$  &
$20$  \\
$\nu_0=5$& $65$ & $47$   &
$17$ \\
\hline
\hline
\multicolumn{4}{|c|}{$d_1=2$}\\
\hline
& $t_f=1$ & $t_f=10$ & $t_f=100$ \\
\hline
$\nu_0=3$& $70$ & $26$  &
$1$  \\
$\nu_0=5$& $43$ & $13$   &
$0.1$ \\
\hline
\end{tabular}
\end{center}\bs{1} \caption{\small Percentage of cases yielding three-dimensional
  decompactification, showing the dependence on $d_1$, $\nu_0$, $t_f$.}\bs{1.7}\label{pertab}
\end{table}

\section{Discussion\label{sect:discussion}}

In this paper we studied a model in which successive thermal
fluctuations are able to decompactify a total of three dimensions.
Although these dimensions are initially anisotropic, it was argued in
\cite{brandisotr} that string gas cosmology will eventually lead to
isotropization.  So roughly speaking our results support the BV
mechanism, as providing a candidate stringy origin of the universe we
observe.  However a more refined statement is that in our model the
probability of decompactifying three dimensions depends on several
parameters, in particular on the fluctuation time $t_f$ and the energy
density $E(t_f)$.  We now discuss the implications of these results
for the BV mechanism.

We found that three large dimensions can arise provided scale factor
fluctuations in the Hagedorn phase occur frequently, on a timescale
$t_f \sim \sqrt{\alpha'}$.  These frequent fluctuations are necessary
so that the dilaton does not have time to roll to weak coupling. It
seems reasonable to take $t_f \sim \sqrt{\alpha'}$, since
$\sqrt{\alpha'}$ is the natural timescale of the system in the
Hagedorn phase.

Assuming $t_f \sim \sqrt{\alpha'}$, the energy density $E(t_f)$ is
basically fixed by the initial value of the dilaton.  To address the
dependence on this quantity consider two successive fluctuations, the
first at time $t=0$, the second at time $t \sim \sqrt{\alpha'}$.  The
initial value of the dilaton $\varphi_0$ determines the subsequent
possibilities.  First, there is a relatively narrow range towards
strong coupling where the equilibrium phase at $t=0$ is
$\mathcal{R}_{d_1}$. In this case we do not expect additional dimensions to decompactify.
So other than invoking anthropic
arguments, we see no way to argue for a three-dimensional outcome. But
if the initial coupling is not too strong, then we start in
$\mathcal{H}$ and the following possibilities remain. One possible
outcome, which fortunately occupies the smallest volume in the space
of initial conditions, is that the system lands in the part of
$\mathcal{R}_3$ where the winding modes wrapping the $d_2$ smaller
dimensions cannot annihilate. In that case, we again expect only $d_1$
dimensions to grow large.  But for most initial conditions the
universe either ends up in the part of $\mathcal{R}_3$ where a total
of three dimensions decompactify, or remains stuck in the Hagedorn
phase.

This latter possibility, of staying in the Hagedorn phase, may seem
discouraging.  But in fact it could work in favor of the BV mechanism,
because roughly speaking it brings us back to square one.  As time
goes by the scale factors may evolve slightly, but if fluctuations
occur frequently enough this evolution is inconsequential. The key
point is that even if the $d_1$ dimensions are unwound and experience
positive pressure, when the system is in equilibrium in $\mathcal{H}$
their growth over many string times is tiny.  This is because the
energy in radiation ($\sim \sqrt{E}$) is much smaller than the energy
in matter ($\sim E$), and furthermore the matter energy is very nearly
independent of $\nu$. This nearly constant large energy keeps the
velocity of the dilaton large, but also results in large ``dilaton
friction'' and a negligible change in $\nu$. This can be seen by
writing the equations of motion in the form
\be\begin{aligned}
\frac{d}{dt}(e^{-\varphi}\dot{\phic})&=-\frac{1}{8\pi^2}E \\
\frac{d}{dt}(e^{-\phic}\dot{\nu})&=\frac{1}{48\pi^2\sqrt{\pi}}\sqrt{E}
\end{aligned}
\ee
In our numerical solutions, even for $t_f = 100$, the most that $\nu$
changed was by one part in a hundred, while the energy remained
constant to one part in a few thousand. Thus the $d_1$ dimensions can
``hover'' in the Hagedorn phase for a long time, which allows
additional fluctuations to take place.  Of course during this time the
dilaton is rolling monotonically towards weak coupling \cite{winds,Danos:2004jz},
so if too much time goes by before another fluctuation takes place,
the string coupling will become so small that the winding modes are
unable to annihilate.  But with a fluctuation timescale $t_f \sim
\sqrt{\alpha'}$, this may not be a significant concern.  To summarize,
{\em within our model we find that there is a considerable window in
which scale factor fluctuations favor the eventual decompactification
of three dimensions, provided such fluctuations occur on timescales}
$\sim \sqrt{\alpha'}$.

We conclude with a few issues which must be addressed in order to
reach a definitive conclusion regarding the robustness of the BV
mechanism.
\begin{itemize}
\item
We found that the BV mechanism can operate provided there are
fluctuations into a regime where winding strings are dilute.  This
raises the crucial issue of whether a dilute string gas in the early
universe can evolve towards a homogeneous cosmology at late times.
Studying this requires going beyond our mini-superspace
approximation, in which we only kept the homogeneous modes of the
metric and dilaton.
\item
We found that the BV mechanism can operate provided fluctuations
in the scale factors occur on timescales of order $\sqrt{\alpha'}$.
It seems reasonable to assume $t_f \sim \sqrt{\alpha'}$ in the
Hagedorn phase.  But this assumption should be validated, for
example by deriving the statistics of fluctuations from a study of
the stochastic evolution of scale factors coupled to a hot string gas.
\end{itemize}

\bigskip
\bigskip
\goodbreak
\centerline{\bf Acknowledgements}
\noindent
The work of DK was supported in part by NSF grants PHY-0855582 and PHY-1214410 and by PSC-CUNY grants.


\providecommand{\href}[2]{#2}\begingroup\raggedright\endgroup

\end{document}